\documentstyle[preprint,aps,prbbib]{revtex}
\draft
\newcommand{\MS}{$\rm MnSb$}
\newcommand{\MA}{$\rm MnAs$}
\newcommand{\MP}{$\rm MnP$}

\begin{document}

\title{Coordination and chemical effects  on the 
structural, electronic  and magnetic properties in 
Mn pnictides}

\author{A. Continenza, S. Picozzi}
\address{Istituto Nazionale di Fisica della Materia (INFM)
Dipartimento di Fisica, \\
Universit\`a degli Studi di L'Aquila, I--67010 Coppito (L'Aquila), 
Italy }
\author{W. T. Geng,  and A. J. Freeman}
\address{Department of Physics and Astronomy and Materials Research Center\\
Northwestern University, Evanston, IL 60208 (U.S.A.)\\}

\maketitle

\begin{abstract}
Simple structures of MnX binary compounds, namely hexagonal NiAs
and zincblende, are studied  as a function of the anion 
(X = Sb, As, P) by means of the
all--electron FLAPW method within  local spin density 
and generalized gradient approximations. 
An accurate analysis of the structural, electronic and magnetic properties
reveals that the cubic structure greatly favours the magnetic alignment
in these compounds leading to  high magnetic moments and
nearly half-metallic behaviour for \MS \ and \MA. The effect of the anion
chemical species is related to both its size and  the possible
hybridization with the Mn $d$ states; both contributions are seen
to hinder the magnitude of the magnetic moment for small and light anions.
Our results are in very good agreement with experiment - where available -
 and show that the generalized gradient approximation is essential
to correctly recover both the equilibrium volume and magnetic moment.

\end{abstract}

\newpage
\section{Introduction}
The renewed interest in diluted magnetic semiconductors \cite{nature,ohnosci}
as promising materials for innovative spin--based devices, is motivating
studies in many related materials. In fact, the addition of the spin degree 
of freedom to
 conventional electronics know--how is paving the way to the appealing  field 
dubbed ``spintronics".
In this work, we  show  results obtained from first principles FLAPW
calculations
for   Mn  compounds.
In particular, we look at the structurally simple binary compounds,
\MS, \MA \ and \MP,  and perform full optimization of the structural
parameters in the hexagonal ferromagnetic NiAs phase,
as well as for the more simple zincblende structure.
We remark that \MP \ is not stable in the NiAs-phase; however, its
equilibrium phase (namely the \MP--type) has a crystal structure which slightly
distorts from the NiAs. Moreover, while \MS \ and \MA \ are both
ferromagnetic in their equilibrium structure, \MP \ shows a complex
antiferromagnetic
helical alignment at low temperatures (below 47K),  is ferromagnetic
between 47 and 291K and paramagnetic at higher temperatures\cite{okuda}. 

The goal of the present work
is to shed light on how {\em i}) 
different coordination
({\em i.e.} distorted octahedra in NiAs--type and tetrahedral in zincblende)
and {\em ii}) different
bonding properties (along the series \MS,  \MA \ and \MP) affect  
the structural, electronic and
magnetic properties of these  Mn compounds.
Moreover, we focus  
on how well the investigated
properties are reproduced within the local spin density (LSD) and 
generalized gradient approximation (GGA).
None of the investigated compounds is stable in the zincblende structure; however,
since Mn  often substitutes for the group III element in III-V compounds,
it is interesting to understand and compare the electronic and
magnetic properties of  Mn in this  coordination.
In addition, it is well established that the zincblende  diluted 
Ga$_{1-x}$Mn$_x$As system is stable for concentrations up to x=7\% and
recent experiments\cite{Abee} on the MBE epitaxial growth of MnSb
(in the NiAs phase) on GaSb reported the presence of another magnetic phase
showing an anomalous Hall effect - attributed to clusters of zincblende 
(Ga,Mn)Sb. 

The paper is organized as follows: in Sect.~\ref{techn} we describe
few details of the calculation;  we discuss results
regarding the structural, electronic and magnetic properties on Sect.~\ref{MS},
\ref{elec} and \ref{magn}, respectively. Finally, we draw
our conclusions in Sect.~\ref{concl}.

\section{Method }
\label{techn}
All  calculations were performed using the all--electron
full--potential linearized augmented plane wave (FLAPW)
method\cite{flapw} within the local spin
density approximation to density functional theory. 
In order to check the accuracy, we performed structural optimizations
using both LSD as parametrized by von Barth and Hedin\cite{LSD} and the
generalized gradient approximation (GGA) as proposed by Perdew and Wang\cite{GGA}.
 We used the following muffin--tin sphere radii   $R_{Mn}, R_{Sb}= 2.40 a.u.$, 
 $R_{As}= 2.0 a.u.$,
 $R_{P}= 1.9 $ a.u.; 
well   converged LAPW wavefunctions 
were achieved using a plane--wave cutoff $k_{\rm max}= 3.0$ a.u.,
resulting in about 300 basis functions (in the  unit cells described below). 
 Inside the muffin--tin spheres, we used an 
 angular momenta expansion up to $l_{max} = 8$
for the potential and  charge density representations; 
the same value   $l_{max} = 8$ was used for the wave--function expansion.
Integrations over the irreducible
Brillouin zone wedge were performed using 24 $k$--points according
to the Monkhorst--Pack scheme\cite{MP}; the accuracy of this mesh
has been checked by comparison with results obtained with  higher
numbers of inequivalent $k-$points.

\section{Structural Results}
\label{MS}

  Our results are summarized
in Table~\ref{nias} and Table~\ref{zincb} for the NiAs (space group
$P_{6_{3}}/mmc-D^4_{6h}$ -- No. 194 in the {\it International Tables for
X-Ray Crystallography}) and the zincblende phase,
respectively, and compared with experiments\cite{Wyc,tesi} and other 
calculations\cite{degroot,sanvito}.
It is clear that  GGA  gives the best agreement 
with experiment (available only for \MS \ and \MA) - recovering in both
cases the correct
volume and shape of the unit cell. 
On the other hand, LSD severely underestimates the equilibrium volumes (by as
much as 20 $\%$), so that its use seems not to be very suitable
 when dealing with equilibrium
properties of Mn pnictides.
The same features have  been 
previously observed for other
compounds containing Mn\cite{terakura} and in the 
more studied case of Fe\cite{Fe}. 
The equilibrium volumes for the
hexagonal structures follow the trend of the anion ionic radii ($R_{Sb}$=1.40
\AA,
$R_{As}$=1.20\AA, $R_{P}$=1.06\AA): we find   about a 20\%  variation 
along the series which leads to about a 7\% variation of the Mn-anion
distance.
However, due to the different $c/a$ ratios, the cell shape 
  also changes: while the volume variaton is considerable,
the value of the $c$ axis of the cell varies only by 4\% from Sb to As
and by 1\%  from As to P. In fact, the vertical
size of the cell seems to be almost fixed by 
the Mn-Mn nearest neighbour distance ($d_{Mn-Mn}$= 2.89, 2.84, and 2.68
\AA \ in \MS, \MA \ and \MP, respectively). Most of the volume variation is 
therefore due to a shrinking of the cell basal plane  which contains the anion (sitting
in the middle of the hexagonal 2-dimensional cell) and 
determines the Mn-anion distance. 

It is useful to recall that in its stable equilibrium phase, \MP \ shows an 
Mn-Mn distance ($d_{Mn-Mn}$= 2.69 \AA\cite{hasegawa}) that is
very close to the one we
found: therefore  the structure
we considered is qualitatively not very far from the equilibrium one. 
Regarding the  magnetic ordering, we note that as expected ferromagnetism is not favoured
at small Mn-Mn distances:  GGA -LMTO calculations\cite{terakura}   
on  different bulk Mn phases showed that for characteristic
Wigner Seitz radii ($r_{WS}$ $\simeq$ 1.4 \AA, corresponding to
$d_{Mn-Mn}$ $\simeq$  2.8 \AA) the most stable alignment is antiferromagnetic; this
might confirm the reason why the ferromagnetic NiAs phase is not
stable for \MP.

Let us now consider the same compounds in the zincblende structure.
We think it is worthwhile  stressing the relevance of this point,
since many of the novel magnetic semiconductors proposed so far
are compounds derived
by their parent non--magnetic semiconductor in this latter phase, in which 
 even large concentrations (up to
50\%\cite{giapp}) of cations have been substituted by Mn. 
As a result, the final coordination
of the Mn atoms is strictly tetrahedral. 

Our results obtained using the GGA approximation are reported on 
Table~\ref{zincb}.
Due to difficulties in stabilizing the cubic  phase, experimental data
for zincblende MnX compounds are not available; however,  
an experimental study \cite{jmmm_ohno}
focused on a Ga$_{1-x}$Mn$_{x}$As system with Mn concentrations up to $x$ =
0.07, showed
 a linear extrapolation at $x$ = 1
of the alloy lattice constant (assuming Vegard's law
validity) leading to $a$ = 5.87 \AA. This differs by 3.7 $\%$ from our
result for MnAs; however, we must remark  that the linear 
extrapolation may indicate
 an incorrect value,
since the actual experimental lattice constant data are reported only 
for very small
Mn concentrations. Moreover, recent calculations \cite{zhao}
pointed out significant deviations from  Vegard's rule in 
Ga$_x$Mn$_{1-x}$As systems.  
Other first principles calculations
of the MnAs lattice constant
performed without GGA give $a$ = 5.66 \AA \cite{sanvito}
and $a$ = 5.87 \AA \cite{shirai_sbag} within a pseudopotential and   
FLAPW formalism, respectively. The differences with our LDA value (not
shown in the tables, $a$ = 5.34 \AA) are probably to
be ascribed to the use of pseudopotential approach in the first case and
to a different minimization approach or FLAPW implementation in the second case.
It needs to be pointed out, however, that a reliable determination of the equilibrium 
lattice constant is particularly important for the corresponding magnetic
properties. In fact, as explained in detail
below,  the magnetic
moment shows a strong dependence on the volume: an overestimate of the
equilibrium zincblende lattice constant may even lead to an incorrect stable
 half-metallic behaviour.  
Furthermore,  from Table~\ref{zincb} we note that
  the equilibrium volume per formula unit ({\em i.e.} per
MnX pair) scales with the anion ionic 
radii in both phases, and it is always larger in the zincblende phase (see also 
Fig.~\ref{struc1}). 
In particular, we remark that while the Mn-anion bond length is very similar
in the NiAs and zincblende phase (only 4\% larger in the cubic phase),
the Mn-Mn nearest neighbour distances are very much larger in the 
zincblende phase ($d_{Mn-Mn}$= 4.36, 3.99, and 3.75
\AA \ for \MS, \MA \ and \MP, respectively). These large
distances favour  ferromagnetic alignment and the direct exchange interaction,
leading to an enhanced magnetic moment.
 It is very interesting to note in Table \ref{zincb} the
very good matching conditions between the MnX zincblende and its Ga--based
analogue: this suggests that under non--equilibrium conditions
Mn may easily occupy the cation sites, if the growth is achieved on suitable
substrates and if   solubility problems are overcome.

In order to show the stability of the different phases, the total energy 
behaviour versus volume/formula--unit is plotted in
Fig.\ref{struc1}. On the same plots we also report the total magnetic moment
as a function of the volume. In all the compounds analyzed, the NiAs--phase is
more stable than the  zincblende phase by more than 0.7 eV per formula unit. 
In addition, the zincblende structure shows a larger equilibrium volume with
respect to the NiAs phase (30\% in MnSb and 25\% in MnP)
joined by a parallel increase in the magnetic moment
per cell. However, the NiAs phase remains the most stable over a large range
of pressures showing that even at the most favorable growing conditions
for  the zincblende phase, the NiAs phase may segregate forming clusters,
as found, for example, in the growth of (Ga,Mn)As at  Mn
 concentrations higher than 7 \% \cite{ohnosci}.

\section{Electronic properties}
\subsection{NiAs phase}
\label{elec}

We now discuss  the electronic properties of the MnX compounds in both phases.
In Fig.\ref{nias-bande}, we plot the band dispersion for the NiAs phase
along the $\Gamma$ -- A direction. 
The band dispersion along this line is very similar in all  three compounds
considered. At high binding energies, we find the anion--$s$ derived bands:
these states are very little affected by the Mn exchange energy and follow
the expected trend in energy  related to the atomic s--levels (5s, 4s, and 3s
for Sb, As and P, respectively). 
At energies closer to the Fermi level ($E_F$), we find the Mn $d$ bands
hybridized with the anion--$p$ states. Actually, the lower states ($\sim$ -5.0
eV)
in this group are those with higher $p$ hybridization (still less
than 7-8 \%), while the others have Mn $d$ character exceeding 
48\% at $\Gamma$. A somewhat larger hybridization is allowed
by symmetry at the A point, reaching about 12\% for  
states well below the Fermi level ($E_F$) for the majority spin component
 ({\it i.e.} states with $A_3$ symmetry bonding
and  non-bonding as indicated in Fig.~\ref{nias-bande} by the
solid and dashed bold lines, respectively). 

The states crossing $E_F$
show high anion $p$ character at  $\Gamma$ (up to 20\%), but become
entirely Mn $d$ states at A. For the minority component we can easily
trace the same features, shifted at lower binding energies due to the exchange
interaction. This latter determines only a partial and small occupation of the lower
non-bonding  $A_3$ split band for {\bf $k$}--vectors very close to the zone center
for \MS \ and \MA, while in \MP \ we find a complete half-filling of the
4--fold degenerate $A_3$ state. 
Looking at the series Sb, As, P, we can see that the states at  $A$ 
denoted by the bold  solid lines in Fig.~\ref{nias-bande} are very close in the
Sb compound while their separation increases as we lower the anion atomic number;
 at the same time we find that the exchange splitting
(i.e. the energy separation between corresponding states for the majority
and minority component) decreases leading to the partial filling of the
higher $A_3$ state in \MP. Moreover, at $\Gamma$ the symmetry allowed $p-d$
 hybridization is seen to increase the band width of the
states originating from the lower $A_3$ state 
(bold solid line in Fig.~\ref{nias-bande}) on going from \MS \ to \MP.
On the other hand, at the $A$  point we find a slightly lower
 anion $p$ component 
 for  \MP \  (7\%) compared to  \MS (12\%), probably  related
to the smaller sphere used for P.

The picture that emerges from the band structure is the following: while
\MS \ and \MA \ show a very similar $p-d$ hybridization, the compound with 
P seems to have a higher hybridization which affects the exchange splitting
of the Mn $d$ states. This is more clearly seen by inspection of the density of
states projected (PDOS) on each atomic site. In Fig.~\ref{DOS}, we show the
PDOS for  \MS \ and \MP \  (in particular Mn $d$ states and anion
$s$ and $p$ states: analogous data for the \MA \ compound
are not reported since they are half-way   between
the two limiting \MS \ and \MP \ cases). In \MP, we found a broader valence band: the P $p$
states start at around -7.5 eV and give rise to two well separated
peaks; the feature at lower binding energies ($\sim$ -3.0 eV) is due to strong
hybridization with the Mn $d$ states that causes a splitting  of the $d$ band.
These two different features can be traced back to the $A_3$ states (see
Fig.\ref{nias-bande}) discussed previously (in \MS \ these states are very close
in energy in the majority component while in \MP \ they are split by about 1.5
eV). The higher hybridization in \MP \ leads to a broadening of the Mn 4$d$ band
joined by a smaller effective exchange energy among $d$ states and a
consequent higher filling of the minority component. 

In order to complete the
discussion of the electronic properties, we compare our calculated GGA
results with available photoemission data and other calculations in 
Tables \ref{relint} and \ref{GA}. The agreement between
experimental\cite{arups} and theoretical
data is noticeably improved by the present GGA calculations in  \MS,
for which there are detailed ARUPS data available; we expect that the same
agreement might hold for the other compounds. 
For  \MA, we compare with UPS spectra\cite{shirai98}
which, however, are not very detailed; experiments find
a broader Mn $d$ band with respect to  \MS \  centered around ~ 2
eV, and this is consistent with our calculation. Also, the tail structure
observed in the \MA \ case at 4-6 eV below $E_F$ is consistent with our
PDOS (not shown) which shows a broadening due to $p-d$ hybridization
in--between  \MS \ and \MP \ case. 

For  \MP, we can compare with recent UPS and IPES data\cite{okuda}
which find (for the real MnP-type structure) a broader
Mn $d$ band compared to Sb,   a small peak at
-0.4 eV and broad structures at -2.6, -6.4 and ~-10 eV. These data
match very well the features found in our calculations and shown in
Fig.\ref{DOS}: very close to $E_F$ we find a small feature related
to Mn $d$ states hybridized with P $p$ states corresponding to
the band crossing $E_F$ and which is very flat at the $A$ point (compare
with Fig.\ref{nias-bande}). At higher binding energies (from $E_F$ down
to 3 eV) the Mn $d$ contribution is predominant 
while the feature at 6 eV, according to our results, has to be ascribed
to P $p$ states. In agreement with our calculations, photoemission experiments
find a much smaller $d$ exchange split energy (i.e. the energy distance between occupied
and unoccupied $d$ states) in \MP \  compared to \MS \ (about 2 eV and more
than 3 eV in \MP \ and \MS, respectively).

We point out that after including  relativistic corrections, namely 
spin--orbit coupling, on  \MS \ (where it is expected to play
the most important role), we found
that the energy splitting for the bands with higher
Mn $p$ content was of the order of 0.5 eV.
However,  the energy levels involved (mainly the Mn $d$
states hybridized with  Sb $p$ states) are at energies far from $E_F$
 and therefore do not affect effectively the magnetic moment or the
exchange interaction among Mn $d$ states, as we will see below.
Therefore, we did not explicitly consider the spin--orbit coupling
on the bands shown, since this would greatly complicate the plots and the
discussion of the band structure without really changing the overall
physics.
\subsection{Zincblende phase}
We now discuss  the electronic properties of these same compounds in the 
zincblende phase. The band structures are reported in Fig.~\ref{zincb-bande}
and the main features can  once again be related to the different degree of
$p-d$ hybridization. As a general characteristic, in agreement with that
already found in similar calculations\cite{degroot}, we find that at $\Gamma$
the lower state is a $3-fold$ degenerate $\Gamma_{15}$ state ($t_{2g}$)
which allows
hybridization with $p$ states, while the doubly degenerate $\Gamma_{12}$ state
($e_g$) has entirely Mn $d$ character. The states with mainly anion $p$ character
are  above $E_F$ and show a very wide dispersion
so that in all three compounds they give rise to electron pockets at the
zone boundaries (namely at X and L). As expected, the exchange interaction
is larger for the non-hybridized $\Gamma_{12}$ states and follows the
same trend with hybridization as previously discussed for the NiAs phase:
 lighter anions have
deeper lying $p$ levels (and are therefore closer in energy to the Mn 3d states) so that
they allow for greater hybridization. In fact, we find that while 
the $p$/$d$ occupation ratio is about 0.06 for \MS, it becomes 0.13 in 
\MP.
This leads to a broader Mn $d$ band and significantly lowers the exchange
interaction along the series, as shown in Table~\ref{GZ}. 

In all cases,
however, the exchange splitting is higher than in the corresponding
NiAs phase showing that the lower coordination 
 greatly favors the $d$-$d$ interaction
leading to a higher magnetic moment, as will be discussed in greater detail below.
The band structures for the minority components show that \MS \ and \MA \ are
nearly half--metallic (the occupied portion of the $\Gamma_{12}$ state is very
close to zero) while  \MP \  shows a much larger occupation factor.
This is of course of great interest in the design of new semiconducting
magnetic materials since it shows that properly grown materials might
present very appealing properties, as hinted by the study of Abe {\it et al.}
on (Ga,Mn)Sb. 

We can compare our results for zincblende \MA \ with a photoemission
study\cite{okabayashi}
performed on cubic $Ga_{1-x}Mn_xAs$. We find that the main features are nicely
reproduced in our calculation for $x$=1; in particular, we find a quite
broad Mn $d$ band centered at around 2.5 eV and  features around
5 eV below $E_F$ that are related to As $p$ -- Mn $d$ hybridization.
Our calculations, however, do not reproduce the satellite state at 6 eV below
$E_F$ that seems to be related to many-body correlation effects. However,
if we look at the exchange splitting for the $\Gamma_{12}$ state (3.4 eV) we
find
that it is remarkably close to the Coulomb repulsion value (U= 3.5 $\pm$ 1.0 eV)
 obtained\cite{okabayashi} by fitting the experimental data with a configuration interaction
(CI) MnAs cluster model.
Moreover, an estimate of the $p-d$ hybridization (the
$pd \sigma$ = 1.0 $\pm$ 0.1 eV parameter fitted in the CI approximation) 
can be obtained from our
calculation by looking at the width of the $\Gamma_{15}$ state at the zone
boundary (~1.16 eV) where the lower split band gains more $p$ weight.
Although our values are calculated in the very high concentration
limit, the agreement with experiment should not be very surprising since
the quantities we have been looking at are really characteristic of the Mn-As
bond and are therefore expected not to be  affected by the  presence of Ga.

\section{Magnetic properties}
\label{magn}
The calculated magnetic properties within GGA are reported in Table \ref{mom} at
the equilibrium volume for both phases considered. The corresponding values
calculated within LSD are not reported, since the equilibrium volumes
are not well reproduced; the LSD calculated magnetic moments at fixed
volume were always smaller (about 5\%) than the corresponding GGA values. 
We should remark that the values in Table~\ref{mom}  take into account
the spin--contribution only; as already mentioned, in fact, spin-orbit
inclusion did not significantly affect  
the total spin magnetic moment (by less than 0.1 $\mu_B$ in the Sb case). 
We also did not consider any orbital
contribution to the total magnetic moment\cite{eriksson}
or the effect of  magnetic anisotropy related to different quantization axes:
the effects have been shown\cite{eriksson} to be of the order of 0.04 and 0.02 $\mu_B$,
for \MS \ and \MA, respectively. 
We are interested in trends and in understanding  how the different
coordination may affect the magnetic properties more than in reproducing 
the exact magnetic behaviour. Moreover, in cubic compounds we expect these
effects to play an even smaller role due to symmetry.

As already pointed out, in all  structures an opposite magnetic alignment
with respect to Mn is present on the anion sites, whose contribution
decreases in going from the hexagonal to the cubic structure.
This characteristic is confirmed by experimental observations\cite{yamaguchi}
 in both \MS \ and
\MA \ and can be traced back to the $p$-$d$ hybridization.
In particular, in \MS \  the total magnetic moment of 3.5 $\mu_B$ with 
a spherically symmetric  magnetic moment
of -0.2  $\mu_B$ at the Sb site is 
reported\cite{yamaguchi}, while in \MA  \ a magnetic moment of 3.4 $\mu_B$ per formula
unit with -0.23  $\mu_B$ on the As sites has been measured by neutron
scattering experiments. The situation in \MP \ is by far more complex due to
the particular structure and the non-collinear magnetism found in this
compound. Looking at the trends along the series
 Sb$\rightarrow$ As $\rightarrow$ P,
Mn shows a higher magnetic moment in the zincblende structure due to the smaller
hybridization with the anion $p$ states in this symmetry; as we mentioned 
earlier, only the $\Gamma_{15}$ state symmetry allows for $p$ hybridization,
while the  $\Gamma_{12}$ symmetry is not compatible with any $p$ character.
This enhances the exchange interaction for these states, whose minority
counterpart is found above $E_F$ and is therefore almost
entirely unoccupied ({\it e.g.} \MS \ and \MA) leading to a nearly
half-metallic behaviour. 

As a result, we find that there are two different mechanisms
that have to be considered when analyzing the trends along the series
Sb, As and P: $i$) a structural effect related to the decreasing anion 
size and $ii$) a chemical effect which changes the state hybridization.
Both contributions result in a  reduction of the magnetic properties
as the anion becomes smaller; in fact, the smaller size leads to a volume
reduction and to a consequent smaller Mn-Mn bond length which increases the
$d$-$d$ overlap and reduces the exchange interaction. At the same time,
the $p$-$d$ hybridization, which is favoured for lighter anions,
also lowers the $d$-$d$ exchange interaction, therefore reducing the resulting
magnetic moment. These same contributions are found in both phases
and play the same role in both coordinations; however,  the zincblende phase
is seen to be the most favorable for enhanced magnetic properties since, due
to symmetry, it matches both requirements: larger volume and reduced 
hybridization.

\section{Conclusions}
\label{concl}

We presented results on the structural and electronic properties of \MS, \MA \
and \MP \ in two different coordinations: hexagonal NiAs  and zincblende.
 Our results show that LSD is not well suited to correctly describe
the structural properties of these compounds since it severely underestimates
the equilibrium volumes. Moreover, at a fixed volume, the calculated spin
contribution to the magnetic moment within LSD is about 5\% smaller than
the one calculated within GGA. 

Comparing our calculated results with available experiments, we found
 very good agreement for the structural properties as well as for the
electronic properties: our findings match very closely
 photoemission data and ARUPS results,
where available. 
Moreover, from a detailed analysis of these compounds in  different phases,
we were able to establish that the cubic coordination greatly favours
the magnetic alignment leading to quite high magnetic moments
due to $i$) lower $p$-$d$ hybridization, and $ii$) larger equilibrium volumes.

On the same footing, we found that in the search for suitable compounds
with high magnetic alignment, the heavier anions should
be preferred since they form compounds with larger volume and
lower $p-d$ hybridization.

\section{Acknowledgements}

Work at Northwestern University supported by the NSF (through the Materials
Research Center).

\begin{table}
\caption[t1]{Calculated structural parameters for \MS, \MA \ and \MP \ 
in the NiAs phase compared with available experiments. The magnetic moment,
$\mu$, is calculated within the muffin-tin sphere. }
\vspace{5mm}
\begin{tabular}{|c|ccc|ccc|ccc|}
 & \multicolumn{3}{c|}{MnSb} &\multicolumn{3}{c|}{MnAs}
 &\multicolumn{3}{c|}{MnP} \\\hline
 &  $a$       &  $c/a$ & $\mu$ &  $a$ & $c/a$ & $\mu$ &  $a$ & $c/a$ & $\mu$ \\
 &    ($\AA$)  &    &($\mu_B$)& ($\AA$)  &  &($\mu_B$)&($\AA$)  &  &($\mu_B$)\\\hline\hline	
this work LSD & 3.863 & 1.37 &  2.46& 3.487 & 1.49 &  1.91 & - & - & -\\
         
this work GGA & 4.128 &  1.35& 3.12 & 3.704 &  1.49& 3.15 & 3.386 &  1.62& 2.04\\

Expt. & 4.120$^a$ -4.13$^b$ & 1.402$^a$-1.396$^b$ & 3.55-3.50 &3.7$^c$ & 1.54 & 3.4$^b$ &-- & --&     \\
\hline
\end{tabular}
\label{nias}
\end{table}

\noindent $^{a}$  Ref. \onlinecite{Wyc}

\noindent $^{b}$    Ref. \onlinecite{tesi}

\noindent $^{c}$    Ref. \onlinecite{sanvito}

\begin{table}
\caption[t2]{Calculated equilibrium lattice constants and magnetic moments for \MS, \MA \ and \MP \ within GGA
for the zincblende phase compared with the experimental lattice constants of
the related zincblende Ga compounds.  }
\vspace{5 mm}
\begin{tabular}{|c|c|c|c|}
 X    & a$_{MnX}$ ($\AA$)& a$_{GaX}$($\AA$)&$\mu_{MnX}  $($\mu_B$) \\ \hline
Sb & 6.166 & 6.096& 3.77 \\
As & 5.643 & 5.657& 3.75 \\
P & 5.308 & 5.451&2.73 \\
\hline
\end{tabular}
\label{zincb}
\end{table}

\begin{table}
\caption[t3]{Calculated exchange splittings (in eV)  and magnetic
moment for MnSb compared with
other calculations and experiments.  }
\vspace{5 mm}
\begin{tabular}{|c|c|c|c|c|}
        &  expt.$^a$ & ASW$^b$ & FLAPW$^c$ & present work GGA\\
\hline\hline	
$\Delta_{ex}$ ($\Gamma$, bond.) & 1.4 $\pm$ 0.3& 1.51 &1.36 &  1.43\\
         
$\Delta_{ex}$ (A, bond.)& 1.7$\pm$ 0.3 &1.72 & 1.56& 1.67 \\

$\Delta_{ex}$ ($\Gamma$, non--bond) & $\sim$ 3.0 & 2.51 & 2.48 & 2.88  \\
$\Delta_{ex}$ (A, non--bond.) & $>$ 2.8 &2.66 & 2.62 &  3.01  \\
$\mu$            & 3.5      & 3.24   &3.21 &3.36 \\
\hline
\end{tabular}
\label{relint}
\end{table}

\noindent $^{a}$ Ref. \onlinecite{arups}.
  
\noindent $^b$ Ref. \onlinecite{degroot}.

\noindent $^{c}$  Ref. \onlinecite{shirai98}.

\begin{table}
\caption[t4]{Calculated exchange splittings (in eV)  and magnetic
moment for \MS, \MA  \ and \MP \ in the NiAs phase.  }
\vspace{5 mm}
\begin{tabular}{|c|c|c|c|c|}
        & $\Delta_{ex} (\Gamma, bond.)$ & $\Delta_{ex} (\Gamma, non-bond)$ 
	 & $\Delta_{ex}$ (A, bond.) &$\Delta_{ex}$ (A, non--bond.)\\
\hline\hline	
\MS \ & 1.43 & 2.88 &1.67 &  3.01 \\
         
\MA \ & 1.397 & 2.87 & 1.60 & 3.02 \\

\MP \ & 0.775 & 1.757 & 0.905 & 1.87  \\\hline
\end{tabular}
\label{GA}
\end{table}

\begin{table}
\caption[t5]{Calculated exchange splittings (in eV)  and magnetic
moment for \MS, \MA  \ and \MP \ in the zincblende phase.  }
\vspace{5 mm}
\begin{tabular}{|c|c|c|}
        & $\Delta_{ex} (\Gamma_{15})$ & $\Delta_{ex} (\Gamma_{12})$ \\
\hline\hline	
\MS \ & 2.01 & 3.61  \\
         
\MA \ & 1.60 & 3.40  \\

\MP \ & 1.16 & 2.83  \\\hline
\end{tabular}
\label{GZ}
\end{table}

\begin{table}
\caption[t4]{Calculated spin magnetic moments (in $\mu_B$) for \MS, \MA \ and \MP \ 
in the NiAs  and zincblende phase  }
\vspace{5 mm}
\begin{tabular}{|c|ccc|ccc|}
 & \multicolumn{3}{c|}{NiAs} &\multicolumn{3}{c|}{zincbl.} \\
 &$\mu_{Mn}$ & $\mu_X$ &  $\mu_T$  & $\mu_{Mn}$ & $\mu_X$& $\mu_T$  \\\hline\hline	
MnSb & 3.41 &  -0.14  &3.26 & 3.86 & -0.09 & 3.77  \\
         
MnAs &  3.18& -0.10  & 3.08 & 3.83&-0.08 & 3.75 \\

MnP & 2.16 & -0.08 & 2.08  & 2.79 & -0.06 & 2.73 \\\hline\hline
\end{tabular}
\label{mom}
\end{table}

\begin{figure}
\caption{Energy versus volume (solid lines) and Mn magnetic moment
(dashed lines) versus volume
 for the MnX compounds (X=Sb, As, P) in the NiAs (circles) and zincblende 
 (squares) phases.  }
\label{struc1}
\end{figure}

\begin{figure}
\caption{Energy bands for the MnX compounds in the NiAs phase along the $\Gamma$--A symmetry
line for the majority (upper panels) and minority (lower panels) spin
components. The  energy zero is set to the Fermi level in each
compound. The thick solid (dashed) lines indicate the bonding (non--bonding)
Mn--$d$ bands.}
\label{nias-bande}
\end{figure}

\begin{figure}
\caption{Energy bands for the MnX compounds in the zincblende phase along the X-$\Gamma$--L symmetry
lines for the majority (upper panels) and minority (lower panels) spin
components. The  energy zero is set to the Fermi level in each
compound.}
\label{zincb-bande}
\end{figure}

\begin{figure}
\caption{Density of states projected on the atomic site for MnSb and MnP
 in the NiAs phase. The  energy zero is set to the Fermi level. }
\label{DOS}
\end{figure}

\end{document}